\documentclass[10pt,a4paper,twocolumn,english,3p]{elsarticle}
\usepackage{mathptmx}

\usepackage[T1]{fontenc}
\usepackage[utf8]{inputenc}
\usepackage{amsmath}
\usepackage{esint}

\makeatletter

\pdfpageheight\paperheight
\pdfpagewidth\paperwidth

\usepackage{babel}
\usepackage{amssymb}
\usepackage{makecell}

\makeatother

\usepackage{babel}
\begin{document}
\begin{frontmatter}
\title{The full strong-coupling expansion of the cusp anomalous dimension
and the $\mathrm{O(6)}$ mass gap}
\author{Zoltan Bajnok$^{a,b}$, Bercel Boldis$^{a,c}$, Dennis le Plat$^{a}$}
\address{\emph{$^{a}$HUN-REN Wigner Research Centre for Physics, 1121 Budapest,
Konkoly-Thege Miklós út 29-33, Hungary$^{2}$}\\
\emph{$^{b}$Institute of Theoretical Physics and Mark Kac Center
for Complex, Systems Research, Jagiellonian University, 30-348 Kraków,
Poland}\\
 \emph{$^{c}$Department of Theoretical Physics, Budapest University
of Technology and Economics, Műegyetem rkp. 3., H-1111 Budapest, Hungary}}
\begin{abstract}
We present the full transseries of the strong-coupling expansion of
the cusp anomalous dimension in $\mathrm{{\cal N}=4}$ super Yang--Mills
theory. This quantity admits an exact representation as a ratio of
two determinants with particularly simple strong-coupling expansions.
Nonperturbative contributions are classified by partitions into distinct
odd integers and obey a universal structure, with Stokes constants
that can be computed iteratively. We also determine the full transseries
of the $\mathrm{O(6)}$ mass gap as a function of the ’t Hooft coupling.
In the AdS/CFT string description, this mass gap fixes the dynamically
generated scale of the O(6) sigma model governing the sphere fluctuations,
and we establish its full nonperturbative relation to the cusp anomalous
dimension.
\end{abstract}
\end{frontmatter}

\section{Introduction}

AdS/CFT duality relates superconformal Yang--Mills theories to string
theories on anti--de Sitter backgrounds \citep{Aharony:1999ti}.
Weak- and strong-coupling descriptions organize the same observables
in very different expansions: perturbation theory on the gauge-theory
side and semiclassical string theory on the dual side. At strong coupling
these expansions are asymptotic and must be completed by nonperturbative
sectors associated with additional saddles. Integrability has provided
the main bridge between these regimes \citep{Beisert:2010jr}.

A central observable in this correspondence is the cusp anomalous
dimension $\Gamma_{\mathrm{cusp}}$ \citep{Belitsky:2006en}. It controls
the ultraviolet divergence of lightlike cusped Wilson loops, the infrared
divergence of gluon scattering amplitudes, and the leading logarithmic
growth of large-spin anomalous dimensions, which are dual to folded
spinning strings. Integrability gives an exact equation for $\Gamma_{\mathrm{cusp}}$
at arbitrary coupling \citep{Beisert:2006ez}. Its notoriously intricate
strong-coupling expansion reveals nonperturbative sectors \citep{Basso:2007wd,Basso:2008tx,Basso:2009gh},
whose first terms were shown to satisfy resurgence relations \citep{Dorigoni:2015dha,Aniceto:2015rua,Dunne:2025wbq}.

The first nonperturbative correction to $\Gamma_{\mathrm{cusp}}$
is closely related to the mass gap of the $\mathrm{O(6)}$ non-linear
sigma model. This model appears in the AdS/CFT string description
on $\mathrm{AdS_{5}\times S^{5}}$: in the appropriate strong-coupling
regime, the sphere fluctuations are governed by an effective $\mathrm{O(6)}$
sigma model \citep{Alday:2007mf}. Although the four-dimensional gauge
theory is conformal, the two-dimensional sigma model dynamically generates
a scale. Its embedding into the full string theory is fixed by a mass
gap $m(\lambda)$, a nontrivial function of the ’t Hooft coupling.
The perturbative part of this mass gap gives, through its square,
the leading nonperturbative correction to the cusp anomalous dimension.

It was further observed in \citep{Dorigoni:2015dha} that $\Gamma_{\mathrm{cusp}}$
admits a representation as a matrix element of the inverse of a semi-infinite
matrix built from Bessel functions. The associated determinant generalizes
the Tracy--Widom distribution \citep{Beccaria:2022ypy,Belitsky:2019fan,Belitsky:2020qir,Belitsky:2020qrm},
and tilted deformations of this structure appear in scattering amplitudes
\citep{Basso:2020xts}.

Recent progress has uncovered the full nonperturbative structure of
these generalized Tracy--Widom distributions and their matrix extensions
\citep{Bajnok:2024epf,Bajnok20251,Bajnok20252,Bajnok:2025lji,Boldis:2026rkb}
together with the full non-perturbative description of the mass-gap
and its tilted generalization {[}{]}. In this Letter we apply these
developments to the cusp anomalous dimension and to the mass gap and
demonstrate further remarkable simplifications.

\section{Transseries for the cusp}

The cusp anomalous dimension appears as the leading logarithmic growth
of large-spin anomalous dimensions. In particular, the anomalous dimension
$\gamma_{S,L}(g)$ of the single-trace operator 
\[
{\cal O}_{S,L}={\rm Tr}(D^{k_{1}}ZD^{k_{2}}Z\dots D^{k_{L}}Z)
\]
 with light-cone derivative $D$, complex scalar $Z$, and total spin
$S=k_{1}+\dots+k_{L}$, exhibits Sudakov scaling at large $S$. This
scaling defines $\Gamma_{\mathrm{cusp}}$ via
\[
\gamma_{S,L}(g)=2\Gamma_{\mathrm{cusp}}(g)\log S+O(S^{0})
\]
 where $g=2\sqrt{\lambda}$ and $\lambda=g^{2}_{Y\!M}N$ is the 't
Hooft coupling and the twist $L$ was kept finite.

The cusp anomalous dimension can be written as a ratio 
\[
\Gamma_{\mathrm{cusp}}(g)=\frac{g}{4\pi}\frac{D_{1}(g)}{D_{0}(g)}
\]
where the quantity $D_{\ell}(g)$ is proportional to a determinant
of a semi-infinite matrix whose matrix elements are given by the kernel
$(e^{\sqrt{x}/2g}-1)^{-1}$ evaluated between appropriately normalized
Bessel functions $J_{n}(x)$, see \citep{Boldis:2026rkb} for details.

The strong coupling expansion of the observable $D_{\ell}(g)$ takes
the form 
\begin{equation}
D_{\ell}(g)=\sum_{\{n_{i}:\mathrm{odd}\}}e^{-n\frac{g}{4}}S^{\{n\}}D^{\{n\}}(g)\label{eq:Dl(g)}
\end{equation}
The summation goes over distinct sets of positive odd integers $\{n\}\equiv\{n_{1},n_{2},\dots,n_{k}\}$
and we use the shorthand $\sum^{k}_{i=1}n_{i}=n$. Thus, at a given
non-perturbative order $n$, multiple contributions appear. The number
of such terms is counted by the fermionic Neveu--Schwartz-type character
$\prod^{\infty}_{j=0}(1+x^{(2j+1)/2})=\sum^{\infty}_{n=0}p_{n}x^{\frac{n}{2}}$.
A distinctive feature of the expansion is that it is linear in each
non-perturbative scale $e^{-(2j+1)\frac{g}{4}}$, in close analogy
with the way fermionic modes enter the character.

Each term $D^{\{n\}}(g)$ is an asymptotic series in $g^{-1}$
\[
D^{\{n\}}(g)=g^{\frac{1}{16}-a^{2}}\sum^{\infty}_{j=0}D^{\{n\}}_{j}g^{-j},
\]
whose coefficients depend on $\ell$; however, for simplicity we suppress
this dependence in the notation. The perturbative sector of $D_{\ell}(g)$
corresponds to $\{n\}=\{\}$. Strikingly, all sectors labelled by
the partitions $\{n\}$ share a universal structure, which takes the
form 
\begin{align*}
D^{\{n\}}_{0} & =1\quad;\quad D^{\{n\}}_{1}=-(a^{2}-\ell^{2})I_{2}\\
D^{\{n\}}_{2} & =(a^{2}-\ell^{2})\left[\frac{1}{2}(a^{2}-\ell^{2}+1)I^{2}_{2}+aI_{3}\right]\\
D^{\{n\}}_{3} & =-\frac{1}{6}(a^{2}-\ell^{2})(a^{2}-\ell^{2}+1)(a^{2}-\ell^{2}+2)I^{3}_{2}+\dots
\end{align*}
The parity of $n_{j}$ modulo $4$ plays an important role and we
define: $\bar{n}_{j}=(-1)^{(n_{j}-1)/2}$. The parameter $a$ depends
on the partition through $\bar{n}=\sum^{k}_{i=1}\bar{n}_{i}$ as $a=\frac{1}{4}-\bar{n}$.
Here $\bar{n}$ measures the difference between the number of terms
of the two parity types. The moments $I_{j}$ also depend on $\{n\}$
as 
\[
I_{j}=J_{j}+2\cdot4^{j-1}\sum^{k}_{i=1}(-1)^{(j-1)\frac{1+\bar{n}_{i}}{2}}n^{1-j}_{i}
\]
where the perturbative coefficients are
\[
J_{2j+1}=2^{2j}\beta(j)\ ;\quad J_{2j}=(2+2^{2j-1}-2^{4j-2})\zeta(2j-1)
\]
Here $\beta(j)$ is the Dirichlet beta function and the term involving
the formally divergent zeta value $\zeta(1)$ is understood in the
regularized sense, yielding $J_{2}=-6\log2$. Observe that the most
transcendental part relevant for QCD is the same at each non-perturbative
order.

The leading $I^{j}_{2}$ behaviour suggests that dependence on $I_{2}$
can be resummed \citep{Bajnok20252}. Indeed, we find that 
\[
D^{\{n\}}(g)=g^{\frac{1}{16}-\ell^{2}}\bar{g}^{\ell^{2}-a^{2}}\sum^{\infty}_{j=0}\bar{D}^{\{n\}}_{j}\bar{g}^{-j}\quad;\quad\bar{g}=g+I_{2}
\]
 where the coefficients $\bar{D}^{\{n\}}_{j}$ are obtained from $D^{\{n\}}_{j}$
by setting $I_{2}$ to zero: $\bar{D}^{\{n\}}_{j}=D^{\{n\}}_{j}(I_{2}\to0)$.
The first coefficients are $\bar{D}^{\{n\}}_{0}=1$ and $\bar{D}^{\{n\}}_{1}=0$.
Higher order terms can be written compactly as $\bar{D}^{\{n\}}_{j}=\frac{(-1)^{j}(a^{2}-\ell^{2})}{j!}\tilde{D}^{\{n\}}_{j}$
with 
\begin{align*}
\tilde{D}^{\{n\}}_{2} & =2aI_{3}\quad;\quad\tilde{D}^{\{n\}}_{3}=2(1+5a^{2}-\ell^{2})I_{4}\\
\tilde{D}^{\{n\}}_{4} & =6(2a^{4}+a^{2}\left(9-2\ell^{2}\right)-\ell^{2}+1)I^{2}_{3}\\
 & \quad+12a\left(7a^{2}-3\ell^{2}+5\right)I_{5}\\
\tilde{D}^{\{n\}}_{5} & =48\left(21a^{4}+7a^{2}\left(5-2\ell^{2}\right)+\ell^{4}-5\ell^{2}+4\right)I_{6}\\
 & +40a\left(5a^{4}+a^{2}\left(37-6\ell^{2}\right)+\ell^{4}-13\ell^{2}+18\right)I_{3}I_{4}
\end{align*}
The physical origin of this structure is not presently understood;
nevertheless, it provides a useful organizational principle that significantly
simplifies the generation and compactification of the expressions.
We provide the first 20 coefficients in an ancillary \emph{Mathematica
}file for $\ell=0,1$. Finally, note that $\bar{g}$ itself depends
on the partition $\{n\}$, and therefore it is not useful to reinterpret
it as a new coupling constant.

The Stokes constant associated with a partition $\{n_{1},\dots,n_{k}\}$
can be generated iteratively. Starting from the normalization $S^{\{\}}=1$
they are obtained using the recursion 
\[
S^{\{n_{1},n_{2},\dots,n_{k}\}}=S^{\{n_{2},\dots,n_{k}\}}S^{\{n_{1}\}}_{a}\prod^{k}_{i=2}\left(1-\frac{n_{1}\bar{n}_{i}}{\bar{n}_{1}n_{i}}\right)^{2\bar{n}_{1}\bar{n}_{i}}
\]
where $a=\frac{1}{4}-\bar{n}$ and 
\begin{align*}
S^{\{n_{1}\}}_{b} & =e^{i\pi(\ell+1+\bar{n}_{1}b)}\frac{\Gamma(\ell-\bar{n}_{1}b)}{\Gamma(1+\ell+\bar{n}_{1}b)}\left(\frac{n_{1}}{4}\right)^{2b\bar{n}_{1}}\times\\
 & \quad\frac{(-1)^{\frac{1}{4}(n_{1}+\bar{n}_{1}+2)}\Gamma\left(1+\frac{n_{1}}{4}\right)\Gamma\left(\frac{1}{2}+\frac{\bar{n}_{1}}{4}\right)}{\Gamma\left(\frac{1}{2}+\frac{n_{1}+\bar{n}_{1}}{4}\right)^{2}\Gamma\left(1-\frac{n_{1}}{4}\right)\Gamma\left(\frac{1}{2}-\frac{\bar{n}_{1}}{4}\right)}
\end{align*}
In particular, for a single element partition one simply has $S^{\{n_{1}\}}=S^{\{n_{1}\}}_{\frac{1}{4}}$.
Higher order Stokes constans can be generated in several different
ways through the recursion, but all constructions lead to the same
final result.

The strong-coupling expansion of $D^{\{n\}}(g)$ leads to factorially
growing coefficients $D^{\{n\}}_{j}$, implying that the transseries
\eqref{eq:Dl(g)} is only formal. To assign it a precise meaning,
it must be understood through lateral Borel resummation \citep{Aniceto:2018bis,Dorigoni:2014hea,Marino:2012zq}.
This procedure replaces the asymptotic series with its Borel transform
\[
B^{\{n\}}(s)=\sum^{\infty}_{j=0}D^{\{n\}}_{j}\frac{s^{j+\frac{1}{16}-a^{2}}}{\Gamma(j+\frac{17}{16}-a^{2})}
\]
which has a finite radius of covergence. Analytic continuation beyond
this domain reveals logarithmic branch-cut singularities on the real
line. The cuts along the positive real line obstruct the naive inverse
Borel transform and one must instead define the lateral Borel resummations
\[
S_{\pm}(D^{\{n\}}(g))=g\int^{\infty e^{\pm i\epsilon}}_{0}e^{-gs}B^{\{n\}}(s)
\]
which differ by an ambiguity originating from the branch cuts. All
these information is encoded in the Stokes autormorphism, which acts
on the transseries and relates the two lateral resummations $S_{+}=S_{-}\circ\mathfrak{S}$.
Its logarithm defines the alien derivatives, which encode the location
of the singularities in the Borel plane as well as the discontinuities
across the cuts. The fact that each non-perturbative scale $e^{-n_{i}\frac{g}{4}}$
appears only ones implies the fermionic-type structure 
\[
\mathfrak{S}=\prod^{\infty}_{j=0}(1+e^{-(2j+1)\frac{g}{4}}\Delta_{2j+1})
\]
which closely resembles the NS-character formula. The final physical
answer is understood as the $S_{+}$ Borel resummation of the transseries
\eqref{eq:Dl(g)}. If instead the $S_{-}$ Borel resummation were
chosen, the transseries would have to be complex conjugated. The action
of the alien derivatives $\Delta_{n_{1}}$ on $D^{\{n\}}(g)$ can
then be computed as 
\begin{align*}
\Delta_{n_{1}}D^{\{n_{2},\dots,n_{k}\}}(g) & =-2i\bar{n}_{1}\sin\frac{\pi}{4}\ \left|\frac{S^{\{n_{1},n_{2},\dots,n_{k}\}}}{S^{\{n_{2},\dots,n_{k}\}}}\right|\times\\
 & \quad\quad D^{\{n_{1},n_{2},\dots,n_{k}\}}(g)
\end{align*}
together with 
\[
\Delta^{2}_{n_{i}}=0\quad;\quad\Delta_{n_{i}}\Delta_{n_{j}}=\Delta_{n_{j}}\Delta_{n_{i}}
\]
We tested these relations against the asymptotic behaviour of the
perturbative coefficients.

One can slightly renormalize the expansion function, such that the
transseries \eqref{eq:Dl(g)} takes the form 
\[
D_{\ell}(g)=\sum_{\{n_{i}:\mathrm{odd}\}}e^{-n\frac{g}{4}}\prod^{k}_{i=1}S^{\{n_{i}\}}\hat{D}^{\{n\}}(g)
\]
This representation allows the introduction of formal parameters $\sigma_{2j+1}$
associated with each non-perturbative scale $e^{-(2j+1)\frac{g}{4}}$.
In this way one obtains a multi-parameter transseries 
\[
D_{\ell}(g,\{\sigma\})=\sum_{\{n_{i}:\mathrm{odd}\}}\left\{ \prod^{k}_{i=1}\sigma_{n_{i}}e^{-\frac{n_{i}}{4}g}\right\} \hat{D}^{\{n\}}(g)
\]
One can then show that the following bridge equation holds 
\[
\Delta_{n_{j}}D_{\ell}(g,\{\sigma\})=-2i\Im m(S^{\{n_{j}\}})\partial_{\sigma_{n_{j}}}D_{\ell}(g,\{\sigma\})
\]
This implies that the Stokes automorphism simply introduces the shift
$\sigma_{n_{j}}\to\sigma_{n_{j}}-2i\Im m(S^{\{n_{j}\}})$. The resulting
prescription is not the median resummation $S_{\mathrm{med}}=S_{+}\circ\mathfrak{S}^{-\frac{1}{2}}$
of the purely perturbative sector $D^{\{\}}(g)$. Instead, at each
non-perturbative order there appears the seed of an additional transseries,
which can be obtained by fixing the transseries parameters as $\sigma_{n_{j}}=\Re e(S^{\{n_{j}\}})=\cos\frac{\pi}{4}\vert S^{\{n_{j}\}}\vert$.

\section{Transseries for the mass gap}

The $\mathrm{O(6)}$ non-linear sigma model enters the large-spin
problem in the generalized scaling limit, where the twist grows logarithmically
with the spin and $j=L/\log S$ is kept fixed. In this regime the
sphere fluctuations dominate the quantum corrections on the string-theory
side \citep{Alday:2007mf}, and the anomalous dimension is described
by the generalized scaling function \citep{Freyhult:2007pz} 
\[
\gamma_{S,L}(g)=(2\Gamma_{\mathrm{cusp}}(g)+\epsilon(g,j))\log S+\dots
\]
It was observed that this function is related to the ground-state
energy density of the $\mathrm{O(6)}$ sigma model, 
\[
\epsilon(g,j)=2(\rho+\epsilon_{\mathrm{O(6)}})\quad;\quad\rho=j/2
\]
 once the $\mathrm{O(6)}$ mass gap $m(g)$ is fixed as a nonperturbative
function of the ’t Hooft coupling \citep{Basso:2008tx}.

The four-dimensional gauge theory is conformal and has no mass gap,
whereas the effective two-dimensional sigma model dynamically generates
a scale. Its embedding into the full $\mathrm{AdS_{5}\times S^{5}}$
string theory fixes this scale as a highly nontrivial function of
the coupling. The resulting strong-coupling expansion is asymptotic
and must be completed by nonperturbative sectors. The tilted generalization
of this mass gap and its full transseries were obtained in {[}{]}.
Here we specialize and simplify that result for the physical case
relevant to the cusp anomalous dimension.

To state the result, recall first the structure of the cusp transseries.
The determinant quantities $D_{\ell}(g)$ have nonperturbative sectors
labelled by finite sets of distinct positive odd integers. However,
$\Gamma_{\mathrm{cusp}}(g)$, being a ratio of two such determinants,
contains repeated sectors after the denominator is expanded. It is
therefore naturally organized by finite multisets, or equivalently
by partitions $\{n\}\equiv\{n_{1},n_{2},\dots,n_{k}\}$ into positive
odd parts 
\[
\Gamma_{\mathrm{cusp}}(g)=\sum_{\{n_{i}:\mathrm{odd}\}}e^{-n\frac{g}{4}}\Gamma^{\{n\}}_{\mathrm{cusp}}(g)
\]
 In the multi-parameter transseries 
\begin{align*}
\Gamma_{\mathrm{cusp}}(g,\{\sigma\}) & =\sum_{\{n_{i}:\mathrm{odd}\}}\left\{ \prod^{k}_{i=1}\sigma_{n_{i}}e^{-\frac{n_{i}}{4}g}\right\} \hat{\Gamma}^{\{n\}}_{\mathrm{cusp}}(g)
\end{align*}
these repetitions are encoded by arbitrary powers of the transseries
parameters $\sigma_{n_{i}}$.

A remarkable feature of the determinant representation of the cusp
anomalous dimension is that the alien-derivative relations for the
determinants can be propagated directly to the transseries of $\Gamma_{\mathrm{cusp}}(g)$.
Acting with $\Delta_{n_{i}}$ on the perturbative ratio $\Gamma^{\{\}}_{\mathrm{cusp}}\sim D^{\{\}}_{1}(g)/D^{\{\}}_{0}(g)$,
and using both the alien algebra and the generating relations for
the nonperturbative sectors of the two determinants, one can recursively
generate all nonperturbative corrections to $\Gamma_{\mathrm{cusp}}(g)$.
This also leads to simple relations among different nonperturbative
sectors, generalizing those first observed for the leading corrections
in \citep{Dorigoni:2015dha}. For a detailed discussion of the strong-coupling
structure of the cusp anomalous dimension, we refer the reader to
\citet{Boldis:2026dcq}.

The mass gap has the same multi-sector structure
\[
m(g)=\sum_{\{n_{i}:\mathrm{odd}\}}e^{-n\frac{g}{4}}m^{\{n\}}(g)
\]
 Its transseries is organized by the same multisets of positive odd
integers, with a sector of total nonperturbative weight $e^{-n\frac{g}{4}}$,
where $n$ is the sum of the parts. A particularly simple relation
is obtained for the square of the mass gap: it is proportional to
the first alien derivative of the cusp anomalous dimension 
\[
m^{2}(g)=4ie^{-\frac{g}{4}}\Delta_{1}\Gamma_{\mathrm{cusp}}(g)
\]
This relation extends the earlier observation that the leading nonperturbative
correction to $\Gamma_{\mathrm{cusp}}$ is given by the square of
the perturbative part of the mass gap.

The determinant structure implies further relations among some sectors;
in particular, the ratios of consecutive pure one-instanton tower
contributions are independent of the level
\[
\frac{m^{\{1^{\otimes(k+1)}\}}}{m^{\{1^{\otimes k}\}}}=\frac{m^{\{1\}}}{m^{\{\}}}
\]
 A similar relation was conjectured in \citep{Dorigoni:2015dha} for
the full nonperturbative structure. We find that this extension does
not hold in general.

Finally, introducing the nonperturbative scale $\Lambda^{2}=e^{-\frac{3i}{4}\pi}\frac{\Gamma\left(\frac{3}{4}\right)}{2\Gamma\left(\frac{5}{4}\right)}\sqrt{g}e^{-\frac{g}{4}}$,
we can write the first few perturbative expansions in the different
nonperturbative sectors explicitly
\begin{align*}
m_{\mathrm{O}(6)}= & \frac{(g)^{\frac{1}{4}}e^{-\frac{g}{8}}}{\Gamma\left(\frac{5}{4}\right)}\left[1+\frac{3-6\log2}{4g}\right.\\
 & -\frac{\Lambda^{2}}{g}\left(1-\frac{15-6\log2}{4g}-\right)\\
\\ & +\frac{\Lambda^{4}}{g^{2}}\left(1-\frac{33-18\log2}{4g}\right)-\\
 & \frac{\Lambda^{6}}{g^{2}}\frac{8\sqrt{\frac{2}{3}}\Gamma\left(\frac{5}{4}\right)^{5}}{\pi\Gamma\left(\frac{3}{4}\right)^{3}}\left(1-\frac{3-6\log2+\sqrt{3}\frac{\Gamma\left(\frac{3}{4}\right)^{1/4}}{\Gamma\left(\frac{5}{4}\right)^{1/4}}}{4g}\right)\\
 & \left.-\frac{\Lambda^{8}}{g^{2}}\frac{2\sqrt{6}\Gamma\left(\frac{5}{4}\right)^{5}}{\pi\Gamma\left(\frac{3}{4}\right)^{3}}\left(1-\frac{5+6\log2}{4g}\right)+\dots\right]
\end{align*}

\section{Conclusions}

In this Letter we derived the complete strong-coupling expansion of
the cusp anomalous dimension. It admits the representation $\Gamma_{\mathrm{cusp}}(g)\approx D_{1}(g)/D_{0}(g)$,
where each $D_{\ell}(g)$ possesses a remarkably simple transseries
structure. The nonperturbative sectors of $D_{\ell}(g)$ are labeled
by partitions into distinct odd integers $\{n_{1},\dots,n_{k}\}$.
For every such partition, the corresponding sector $D^{\{n\}}(g)$
takes a universal form when expressed in terms of its associated moments
$I_{j}$ and the Stokes constants can be computed iteratively. This
universality indicates that semiclassical expansions of the string
path integral around the different saddle points contributing to $D_{\ell}(g)$
are related in a simple and structured manner. It would be desirable
to understand the underlying string-theoretic origin of this property.

The cusp anomalous dimension governs the large-spin limit of higher-twist
operators, while subleading corrections in the generalized scaling
limit are controlled by the $\mathrm{O(6)}$ sigma model \citep{Alday:2007mf,Basso:2007wd,Basso:2008tx,Basso:2009gh,Bajnok:2008it}.
The embedding of this effective sigma model into the full string theory
is fixed by the $\mathrm{O(6)}$ mass gap, viewed as a nonperturbative
function of the ’t Hooft coupling. Using the nonperturbative relation
between the square of the mass gap and the cusp anomalous dimension,
we obtained the full transseries of the mass gap. Together with the
recently identified transseries structure of the $\mathrm{O(6)}$
model \citep{Bajnok:2022xgx,Bajnok:2025mxi}, this opens the way to
a complete resurgent description of the high-spin regime.

The dressing phase of ${\cal N}=4$ SYM theory \citep{Beisert:2006ez}
is likewise related to matrix elements of the $K$-matrix mentioned
above. Clarifying how the resurgence properties of the dressing phase
\citep{Arutyunov:2016etw} connect to the structure uncovered here
would be an interesting direction for future investigation.

\section{Acknowledgements}

We thank Gregory Korchemsky for reading the draft and the useful comments.
The research was supported by the Doctoral Excellence Fellowship Programme
funded by the National Research Development and Innovation Fund of
the Ministry of Culture and Innovation and the Budapest University
of Technology and Economics, under a grant agreement with the National
Research, Development and Innovation Office (NKFIH). ZB was supported
in part by a Priority Research Area DigiWorld grant under the Strategic
Programme Excellence Initiative at the Jagiellonian University (Kraków,
Poland). DlP acknowledges support by the Alexander von Humboldt Foundation.
The research was supported also by the grant NKKP Advanced 152467.

\bibliographystyle{unsrt}
\bibliography{ref}

\end{document}